# Do Cloaked Objects Really Scatter Less?

Francesco Monticone and Andrea Alù[*]

Department of Electrical and Computer Engineering, The University of Texas at Austin,

1 University Station C0803, Austin, Texas 78712, USA

[*]alu@mail.utexas.edu

*We discuss the global scattering response of invisibility cloaks over the entire frequency spectrum, from static to very high frequencies. Based on linearity, causality and energy conservation we show that the total extinction and scattering, integrated over all wavelengths, of any linear, passive, causal and non-diamagnetic cloak necessarily increases compared to the uncloaked case. In light of this general principle, we provide a quantitative measure to compare the global performance of different cloaking techniques and we discuss solutions to minimize the global scattering signature of an object using thin, superconducting shells. Our results provide important physical insights on how invisibility cloaks operate and affect the global scattering of an object, suggesting ways to defeat countermeasures aimed at detecting cloaked objects using short impinging pulses.*

## I. INTRODUCTION

The last decade has witnessed a significant progress in the field of classical electrodynamic theory and applications, following the introduction and development of the concept of metamaterials. This field of research has provided a new theoretical framework and practical tools to control electromagnetic waves and fields in unprecedented ways, *beyond* what is directly offered by natural materials [1]. Among a broad landscape of fascinating and innovative applications, invisibility devices, or cloaks, are arguably the quintessential example of how metamaterials can control and engineer wave propagation in unconventional ways. The seminal papers on plasmonic [2] and transformation-optics [3] cloaking demonstrated that a properly designed metamaterial shell wrapped around a given object can drastically suppress its scattering for any angle of incidence and observation, making it almost completely undetectable, i.e., invisible, at the frequency of interest. These concepts have had a large scientific impact and keep triggering considerable media interest, especially after their experimental verification at microwave frequencies [4]-[5], making the topic of cloaking one of the major driving forces in the field of metamaterials. Several other techniques to achieve strong scattering suppression in different regions of the electromagnetic spectrum have been investigated (some other notable examples are the mantle cloaking [6]-[7] and waveguide cloaking [8], which have both been experimentally verified). The cloaking concept has also been translated to the acoustic and quantum domains, realizing acoustic [9]-[11] and matter-wave cloaks [12]-[14].

The vast amount of theoretical and experimental results on the topic allows identifying the main challenges associated with this technology, namely, the detrimental effect of ohmic losses and the narrow operational bandwidth, factors that affect in a way or another all mentioned methods. Although bandwidth issues are generally understood to be associated with fundamental causality limitations, independent of the cloak design [15]-[20], recent papers have made important claims of invisibility over very broad bandwidths [21]-[23]. We may say that these solutions are based on smart optical engineering, aimed at rerouting the impinging beams from specific directions with properly arranged mirrors or dielectric lenses, allowing the transfer of an image from the back to the front of an object for specific observers' positions. Although interesting engineering advances, these solutions should not be considered in the same class as the previously mentioned approaches to cloaking, since they do not aim at cancelling the total scattering of an object, but rather they introduce a large, frequency-dependent phase delay to the wave propagating through the designed optical system. This makes the optical trick easily detectable by moving the position of observation or measuring the interference with the impinging beam.

In this paper, on the contrary, we analyze the performance of cloaking devices in restoring the impinging field distribution all around a given object by suppressing its total scattering cross section (SCS). Our ultimate goal is to evaluate the performance of an arbitrary cloaking shell in minimizing the difference between total and incident fields around the object that it surrounds:

$$\int_{\infty/object} |E_{inc} - E_{tot}| dV , \qquad (1)$$

where $\infty/object$ indicates the entire space except the cloaked object. If the integrated difference between total and incident fields in (1) can be drastically reduced, the object becomes truly undetectable, for any observer's position. Most of the works on cloaking [defined as in (1)] have so far focused on suppressing the scattering at a single frequency, and only a limited number of recent papers have investigated the bandwidth performance around the operating frequency (see, e.g., [15]-[20] and [24]-[26]). However, if a cloak is able to reduce the scattering only over a narrow frequency range, it may be easily defeated by sending a short pulse carrying a broad frequency spectrum. It is relevant to assess, therefore, what happens in other regions of the electromagnetic spectrum when an invisibility cloak is wrapped around a given object, and whether one may come up with an optimal strategy to minimize the visibility of an object over a very broad range of frequencies. In the following, we address these questions and define global bounds on cloaking over the entire spectrum. This problem is not only of relevant scientific significance, but it has also crucial practical and technological implications. In many scenarios, from electromagnetic compatibility to warfare, it is of vital importance to understand whether an object cloaked at a given frequency may become a "beacon" in a different range of frequencies.

## II. CLOAKED OBJECTS ACTUALLY SCATTER MORE

It may be qualitatively appreciated studying the many examples of cloaks available in the literature that, while the scattering cross section can be suppressed at a given frequency with many different techniques, we regularly end up "paying the price" with increased scattering outside the narrow cloaking bandwidth. For example, we recently discussed in [27] how in the plasmonic cloaking technique each scattering zero necessarily comes in pair with a scattering resonance at a different frequency. To quantitatively assess how a generic cloak *globally* affects the scattering response, our first step involves using the fundamental sum rule on the extinction cross section of an arbitrary object first introduced by Purcell [28]. By simply assuming linearity, causality and energy conservation [29], it is possible to write in generalized form [30]

$$I_{ext} = \int_0^\infty C_{ext}(\lambda) \, d\lambda = \pi^2 \left( \hat{p}_e^* \cdot \boldsymbol{\alpha}_{e,s} \cdot \hat{p}_e + \hat{p}_m^* \cdot \boldsymbol{\alpha}_{m,s} \cdot \hat{p}_m \right), \qquad (2)$$

which directly relates the extinction cross section $C_{ext}(\lambda)$, integrated over the whole wavelength spectrum, to the static electric and magnetic polarizability dyadics $\boldsymbol{\alpha}_{e,s}$ and $\boldsymbol{\alpha}_{m,s}$; $\hat{p}_e$ and $\hat{p}_m = \hat{k} \times \hat{p}_e$ are the polarization and cross-polarization unit vectors, respectively, and $\hat{k}$ denotes the incident direction. For simplicity and clarity, we concentrate in the following on the lossless scenario, for which the extinction cross section $C_{ext}(\lambda)$ is equal to the scattering cross section $C_{scat}(\lambda)$, but similar considerations may be extended to lossy objects [31]. Eq. (2) shows that, although the local frequency dispersion may be very complicated, on a global scale any linear, passive and causal object scatters an amount of power directly related to the properties of the object at zero frequency. In the case of a homogenous and isotropic sphere, which is a common benchmark example for cloaking, Eq. (2) reduces to

$$I_{scat} = \int_0^\infty C_{scat}(\lambda) \, d\lambda = \pi^2 \left( \alpha_{e,s} + \alpha_{m,s} \right) = 3\pi^2 V \left( \frac{\varepsilon(0)-1}{\varepsilon(0)+2} + \frac{\mu(0)-1}{\mu(0)+2} \right), \quad (3)$$

where the static polarizabilities are scalar quantities; $V = 4\pi a^3/3$ is the sphere volume, $a$ its radius, and $\varepsilon(0)$, $\mu(0)$ are its static relative permittivity and permeability. This explicitly shows that the integrated scattering of a sphere grows when its volume or static constitutive parameters increase.

In light of these sum rules, we can analyze the effect of adding a cloak around an arbitrary object: Fig. 1 shows the normalized SCS over a broad wavelength range for a dielectric sphere with relative permittivity $\varepsilon = 5$ (panel a) and a perfectly conducting (PEC) sphere (panel b), both with diameter $2a = 100 \text{ nm}$. The spheres are then covered with different cloaks: a plasmonic cloak (blue curves), a mantle cloak (red) and a transformation-optics cloak (green), all designed to suppress the SCS at the central wavelength $\lambda_c = 300 \text{ nm}$, as detailed in the following sections.

## A. Uncloaked particles

The scattering of the uncloaked spheres already shows some interesting features: it appears that the dielectric sphere globally scatters more, over the considered spectrum, than a PEC sphere with same radius, while intuition would suggest that an impenetrable object affects more strongly the impinging wave, and thereby should scatter more energy. The sum rule (3) allows addressing this issue by simply comparing the static polarizabilities: a dielectric sphere has a purely electric response ($\alpha_{m,s} = 0$), while a PEC sphere is characterized by both magnetic and electric polarization, since conductors are characterized by static diamagnetism. When an external static magnetic field is applied on a conductor, in fact, its free electrons circulate in eddy currents [32]-[33]. In a PEC this effect produces an ideal diamagnetic response, producing a zero magnetic permeability $\mu(0) \to 0$. At the same time the electric permittivity of a metal has a pole at zero frequency due to its finite conductivity, and $\varepsilon(0) \to -\infty$. The two effects have opposite contributions in (3). The ratio of static polarizabilities between the two spheres is therefore

$$r_\alpha = \frac{\alpha_{s,\text{diel}}}{\alpha_{s,\text{PEC}}} = \frac{4\pi a^3 \dfrac{\varepsilon(0)-1}{\varepsilon(0)+2}}{4\pi a^3 \dfrac{1}{2}} = 2\frac{\varepsilon(0)-1}{\varepsilon(0)+2}, \tag{4}$$

which is larger than unity for $\varepsilon(0) > 4$. In our case $\varepsilon = 5$, which yields an integrated scattering $I_{scat} = 8.86 \cdot 10^{-3} \, \mu\text{m}^3$ calculated through Eq. (3), which is larger than the integrated scattering of the PEC object $I_{scat,PEC} = 2\pi^3 a^3 = 7.75 \cdot 10^{-3} \, \mu\text{m}^3$. These numbers may be confirmed by numerically integrating the curves in Figs. 1a and 1b. Further physical insights on these values is gained by qualitatively comparing the curves for uncloaked spheres in Figs. 1a and 1b: notably, the scattering of the PEC particle exhibits a longer "tail" at long wavelengths due to the excitation of both electric and magnetic dipolar modes in the quasi-static regime. This is compensated in the dielectric case by much larger scattering peaks in the dynamic regime, in particular for wavelengths comparable to the size of the particle.

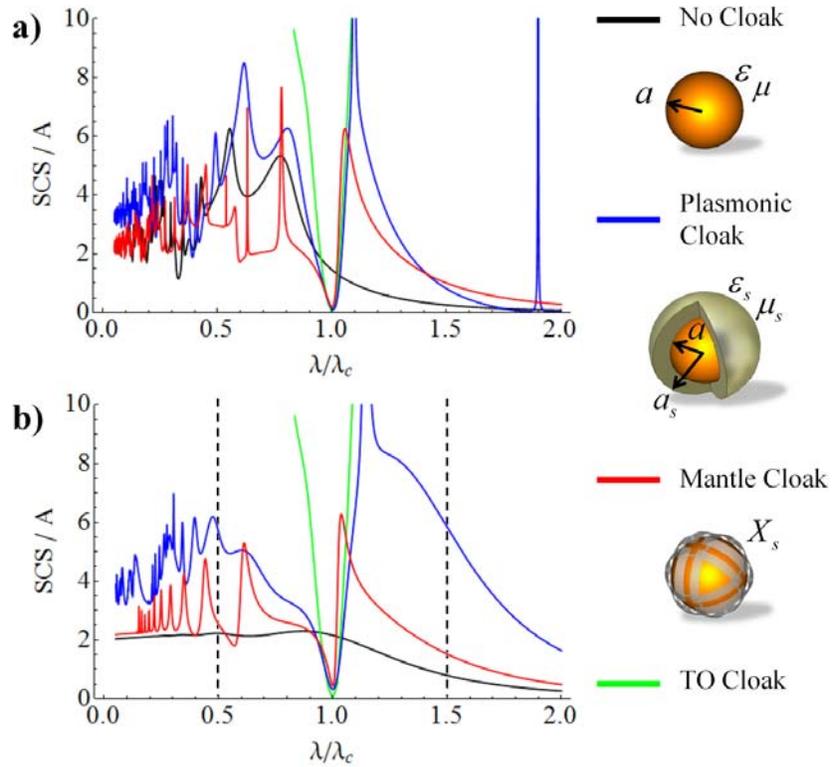

*Figure 1 – Normalized scattering cross section over a broad range of wavelengths for different cloaking techniques: the normalization factor A is the geometrical cross section of the uncloaked sphere. The cloaks are designed to suppress the scattering of: (a) a dielectric sphere with $\varepsilon = 5$ and diameter $2a = 100$ nm, (b) a PEC sphere of same size, at the central wavelength $\lambda_c = 300$ nm. In both panels, the scattering spectrum of the uncloaked objects (black curves) is compared to the response of different cloaking designs: plasmonic cloak (blue), mantle cloak (red) and transformation-optics cloak (green). The vertical dashed lines in (b) enclose the bandwidth considered for the time-domain animations in Figs. 2c,d.*

In Fig. 1 it is also observed that at very short wavelengths the scattering cross sections converge to twice the sphere geometrical cross section $A$ (projected area seen by the impinging wave) independent of the material properties of the object, a general result known as 'extinction paradox' [29]. This implies that an impenetrable sphere and a transparent sphere (say with $\varepsilon =1.01$) with same geometrical cross-section, much larger than the wavelength, exhibit the same scattering cross section, equal to $2A$. This apparent paradox is associated with the fact that, even if the wave can penetrate the transparent sphere, it still accumulates a significant phase delay propagating through the electrically large object. The phase difference with the impinging beam produces an equally large scattering cross section according to Eq. (1), offering a good example of why transparency or the possibility of seeing behind a large object, as in [21]-[23], are not necessarily synonyms of limited scattering. The phase variation behind the transparent sphere may be easily detected with interferometric techniques.

B. Cloaked particles

When the dielectric and PEC spheres analyzed in the previous section are covered by a cloak designed to suppress the SCS at $\lambda_c = 300$ nm, we can qualitatively see that all examples shown in Figs. 1a and 1b are successful in reducing the scattering to very low levels at the design frequency, but they significantly increase it in different regions of the electromagnetic spectrum. The first cloak we have considered in Fig. 1 is a plasmonic cloak (blue lines), realized with a homogeneous spherical shell with low permittivity tailored to cancel the dominant portion of the scattered wave [2]. As in the uncloaked case, the sum of electric and magnetic static polarizabilities determines the integrated scattering, yielding

$$I_{scat} = 4\pi^3 a_s^3 \left[ \frac{(\varepsilon_s - 1)(\varepsilon + 2\varepsilon_s) + \eta^3 (\varepsilon - \varepsilon_s)(1 + 2\varepsilon_s)}{(\varepsilon_s + 2)(\varepsilon + 2\varepsilon_s) + 2\eta^3 (\varepsilon - \varepsilon_s)(\varepsilon_s - 1)} + \frac{(\mu_s - 1)(\mu + 2\mu_s) + \eta^3 (\mu - \mu_s)(1 + 2\mu_s)}{(\mu_s + 2)(\mu + 2\mu_s) + 2\eta^3 (\mu - \mu_s)(\mu_s - 1)} \right], (5)$$

where $\varepsilon$ and $\varepsilon_s$ ($\mu$ and $\mu_s$) are the static relative permittivity (permeability) of the core and shell, respectively, and $\eta = a/a_s$ is the ratio of the core radius $a$ to the shell radius $a_s$. For the plasmonic shell, we assumed in Fig. 1 a lossless silver permittivity, which follows a (causal) Drude dispersion $\varepsilon_s = \varepsilon_\infty - f_p^2/f^2$, with $\varepsilon_\infty = 5$ and plasma frequency $f_p = 2175$ THz [35]. By using the design formulas in [2], the optimal aspect ratio is found to be $\eta = 0.82$ for the dielectric and $\eta = 0.78$ for the PEC case.

At low frequencies, the permittivity of silver becomes very large and negative, i.e., $\varepsilon_s \to -\infty$, and we can neglect its weak diamagnetism [32],[36] letting $\mu_s = 1$. If the core is dielectric, with static material properties $\varepsilon = 5$ and $\mu = 1$, the integrated scattering of the cloaked sphere yields $I_{scat} = 2.79 \cdot 10^{-2} \mu m^3$, which is 3.15 times larger than the uncloaked case. This clearly shows that the addition of a plasmonic cloak actually increases the integrated scattered power. When the cloaked sphere is made of PEC, the integrated scattering (5) simplifies to

$$I_{scat} = \int_0^\infty C_{scat}(\lambda) d\lambda = 2\pi^3 a_s^3 \left(2 - \eta^3\right), \tag{6}$$

which gives $I_{scat} = 2.49 \cdot 10^{-2} \mu m^3$. This is again 3.21 times larger than the uncloaked case.

Another recently proposed technique to suppress the SCS is based on 'mantle cloaking' [6]-[7], which relies on inducing a suitable current distribution on a patterned metasurface (such as a frequency-selective surface [37]) surrounding the object, aimed at canceling the scattering in all directions. Since this technique is not based on volumetric material properties, we may expect that it will affect less the global scattering response. Following the full-wave design procedure in [6], we find that the SCS of the dielectric sphere can be minimized at $\lambda_c = 300$ nm by using a conformal metasurface (aspect ratio $\eta = 1$) with inductive surface reactance $X_s = \omega L = 56.2$ Ω at the design frequency. For the PEC sphere, cloaking is obtained with $\eta = 0.91$, defined as the ratio of the core radius over the metasurface radius (the shunt impedance of the metasurface cannot be directly in contact with a conducting object to avoid a short circuit) and the required reactance is found to be $X_s = \omega L = 26.3$ Ω at the design frequency, when assuming an air gap between the core and the mantle cloak. In both cases we assumed a (causal) linear dispersion for the surface impedance versus the angular frequency $\omega$, according to which at very low frequencies $X_s \to 0$ and the impinging wave sees a conducting, perfectly reflecting spherical surface. The integrated scattering becomes $I_{scat} = 4\pi^3 a_s^3$, according to Eq. (3), which gives $1.55 \cdot 10^{-2}\,\mu m^3$ in the dielectric case and $2.06 \cdot 10^{-2}\,\mu m^3$ in the PEC case. This is consistent with the results in Fig. 1 (red lines), which visually show that mantle cloaks globally scatter less than plasmonic cloaks, but still largely increase the integrated scattering compared to the uncloaked case: 1.75 and 2.65 times larger than the uncloaked dielectric and PEC cases, respectively.

Figs. 1a and 1b also show the case of a transformation-optics cloak, which is based on a suitably tailored inhomogeneous and anisotropic metamaterial shell able to bend the impinging wave around its core region, making any object placed in this area undetectable to arbitrary outside observers [3]. Since the fields do not reach the cloaked region, the cloak response is independent of the covered object and therefore the same curve is obtained in Figs. 1a and 1b. We calculated the SCS curves of Fig. 1 (green lines) using full-wave numerical simulations [38], considering the ideal parameters of the cloak as in [3] for a radius twice as large as the covered spheres, and a realistic (causal) frequency dispersion [39] (an alternative approach based on a large number of homogenous layers with dispersive properties was proposed in [24]). Although in principle a transformation-based cloak can have an arbitrarily small thickness, any practical realization requires an overall size significantly larger than the object to be cloaked, and in our design we considered an outer radius twice as large as the radius of the core, as in [5]. Since the numerical simulation of such structure for short and long wavelengths compared to the size of the cloak poses serious numerical challenges, and an analytical solution is not available for arbitrary frequencies, we show in the figure the simulated cloak response over a relatively narrow bandwidth around the design frequency. Nevertheless, the results at hand show that, while this cloak is the most successful of the three considered here in bringing the scattering to zero at the design frequency, the SCS rapidly increases above the level of the uncloaked spheres, consistent and even higher than the previous cases. Using Eq. (2), and assuming $\varepsilon_s \to -\infty$ and $\mu_s = 2$ in the static limit, consistent with the dispersion models used in our numerical simulations [39], we find $I_{scat} = 15.5 \cdot 10^{-2} \mu m^3$ for this example, which is significantly larger than all the previous examples, and 17.5 and 20 times larger than the uncloaked dielectric and conducting spheres, respectively.

To further confirm these results and help visualizing how a cloak affects the scattering over a broad bandwidth, we performed full-wave time-domain simulations [40] of the PEC sphere and the corresponding plasmonic cloak considered above. When the excitation is a monochromatic plane wave with wavelength $\lambda_c = 300$ nm, the time snapshots of the electric field shown in Figs. 2a,b confirm that the plasmonic cloak significantly reduces the scattering compared to the uncloaked case, and the propagating wavefront is almost unperturbed by the cloaked sphere.

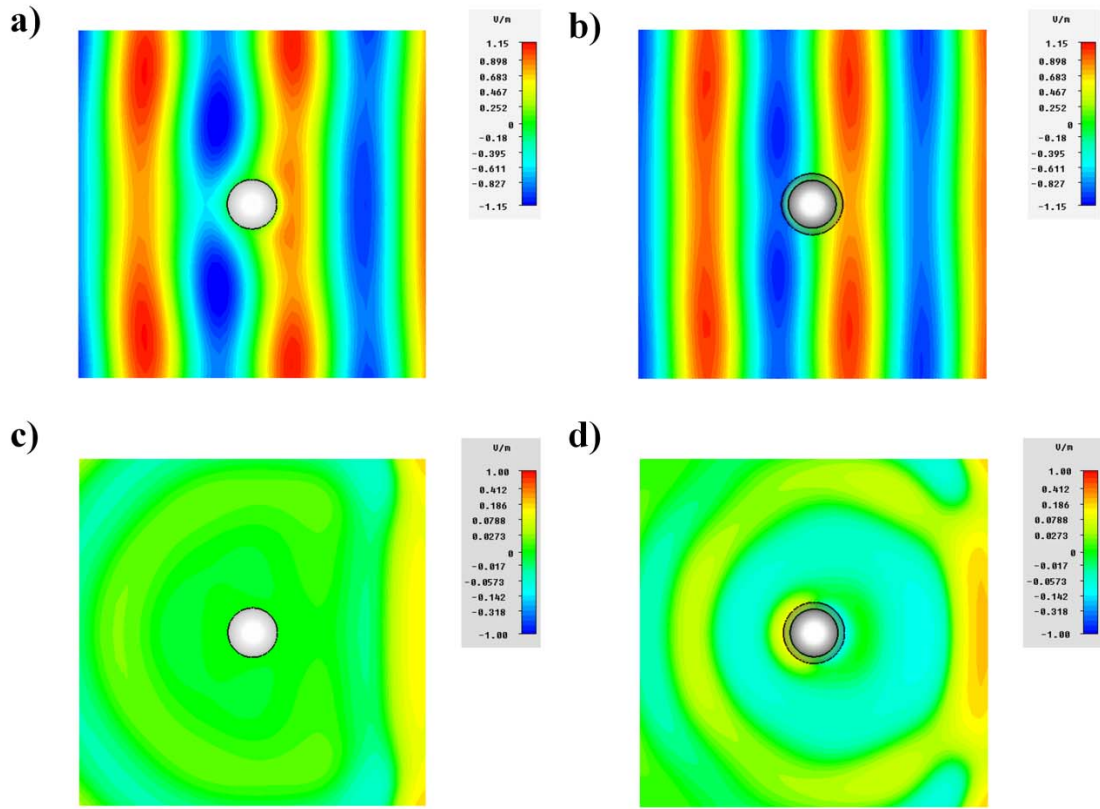

*Figure 2 – Top row: Time snapshot of the electric field distribution on the H plane for monochromatic plane wave excitation at the design wavelength $\lambda_c = 300$ nm. (a) Uncloaked PEC sphere and (b) PEC sphere covered by a plasmonic cloak. Bottom row: Time snapshot of the electric field distribution on the H plane for a Gaussian-pulse plane wave excitation with bandwidth indicated by the vertical dashed line in Fig. 1b, after having just passed the object. (c) Uncloaked PEC sphere and (d) PEC sphere covered by a plasmonic cloak. The geometry of the sphere and the cloak are the same as in Fig. 1b.*

The situation reverses when the object is illuminated by a short plane wave pulse in time, with a bandwidth indicated by the vertical dashed lines in Fig. 1b, over which the integrated SCS of the cloaked sphere is generally larger than that of the uncloaked object. The time snapshots in Figs.

2c,d confirm that when the main portion of the pulse passes the object, the cloaked sphere scatters drastically more than the uncloaked one, due to the frequency dispersion of the cloak material.

Overall, the scattered power from the cloaked object is significantly larger than the uncloaked case for this pulsed excitation, and the cloaked object may be actually more easily detected. In [41] we present the corresponding time-domain animations, which, consistent with this discussion, show the response of the cloaked and uncloaked objects under monochromatic and pulsed excitations. For the given choice of pulse shape, the animations are very effective at showing how the cloak can significantly suppress the scattering from the sphere while the pulse field distribution covers the whole volume of the object. At this moment in time, the cloak mostly feels the central frequency of the pulse, at which it was designed to operate. However, in the transient regimes when the pulse hits or has left the object, the cloak is seen to scatter a significant amount of power, due to its dispersive nature and resonant response, making the cloaked object significantly more visible than the uncloaked one.

## C. Monotonicity theorem and global bounds on cloaking

The results in the previous section suggest that conventional cloaking techniques based on passive and causal covers tend to increase the global scattering integrated over the entire electromagnetic spectrum. In this section we prove that this is indeed true for most cloaking schemes, and discuss the conditions under which a cloak may be designed to instead reduce the total integrated scattering from a given object. To address this point, we use the variational results originally derived in [42], which demonstrate the monotonicity of the static electric and magnetic polarizabilities with the local permittivity and permeability. This powerful theorem proves that, given an arbitrary volume $V$, an increase (decrease) in the static $\varepsilon$ or $\mu$ in any portion of this volume is directly associated to an increase (decrease) in its static electric or magnetic polarizability. As a corollary of this general result applied to our problem, we consider the volume of an arbitrary cloaked object. Since the difference between cloaked and uncloaked cases resides in the addition of a cloak, the electric (magnetic) static polarizability in the presence of the cloak is necessarily larger than in the case without if, within the cloak volume, the local $\varepsilon(0)$, $\mu(0)$ are in all points larger than unity (relative to the background material, considered free-space here). This monotonicity theorem applies to arbitrary isotropic [42] and anisotropic materials [43], essentially covering any cloaking technique available to date.

Next, we note that the static electric permittivity of any linear and causal material obeys the following sum rule [29], directly derived from Kramers-Kronig relations,

$$\mathrm{Re}\left[\varepsilon(0)\right] = 1 + \frac{2}{\pi} \int_0^\infty \frac{\mathrm{Im}\left[\varepsilon(\Omega)\right]}{\Omega} \, d\Omega. \tag{7}$$

Therefore, if we assume that the cloak is strictly passive, $\mathrm{Im}\left[\varepsilon(\Omega)\right] \geq 0$, the real part of the static permittivity at any point in the cloak volume is constrained to be larger than the free-space permittivity [44]

$$\mathrm{Re}\left[\varepsilon(0)\right] > 1. \tag{8}$$

This, combined with the previous discussion, confirms that any linear, causal, non-diamagnetic and passive cloak necessarily increases the overall static polarizability of the object. Notice the requirement of non-diamagnetism, since, as discussed more extensively in the following section, the static permeability does not necessarily follow a condition similar to Eq. (8). Invoking also the sum rule (2), we can therefore state the following theorem, which confirms the observations in the previous section: *any cloak made of linear, passive and non-diamagnetic materials always increases the integrated scattering of the original uncloaked object. This defines a global bound for cloaking, as no cloak with the above properties can do better, in terms of global scattering, than the original uncloaked object*. Suppressing the scattering in some frequency window has to be paid back, with interests, in the rest of the spectrum. In addition, the larger is the cloaking bandwidth around the central frequency, the more scattering in the rest of the spectrum is expected, and therefore any cloak with the above properties is more detectable with a short, broadband pulsed excitation than the original object. There are other fundamental constraints, beyond the global bound derived here, that limit the continuous bandwidth over which scattering may be minimized [20], but this goes beyond the scope of this paper and will be discussed more extensively in future works.

The present finding is particularly powerful as the great majority of cloaks proposed in the recent literature falls within the assumptions of the above theorem, namely, linearity, passivity and absence of noticeable static diamagnetism. Our findings demonstrate that basically all the existing invisibility cloaks inherently increase the integrated scattering of the objects on which they are applied.

## III. DIAMAGNETIC AND SUPERCONDUCTING CLOAKS FOR GLOBAL SCATTERING REDUCTION

After having demonstrated that the integrated scattering of conventional cloaks is bounded, it is interesting to explore whether it is physically possible to do better with alternative designs inspired by these findings. In this regard, we note that a sum rule analogous to (7) for the static magnetic permeability is not available [45], because the function $\mu(\omega)$ loses its physical meaning at moderately high frequencies, before converging to unity. For this reason the previously derived theorem holds an exception for materials with static diamagnetism $0 < \mu(0) < 1$, a property commonly achieved in various natural and artificial materials. The monotonicity theorem implies that the static magnetic polarizability of a body surrounded by a static diamagnetic material is actually lower than the one without cover, hence lowering the magnetic contribution to the integrated scattering. However, in dielectric diamagnetic materials the real part of the static refractive index is required to be larger than unity to ensure causality [45], i.e., $n(0) = \sqrt{\varepsilon(0)\mu(0)} > 1$, and therefore the static electric permittivity $\varepsilon(0)$ must necessarily increase to balance a lower-than-unity $\mu(0)$ [46], leading to the condition $\varepsilon(0) > 1/\mu(0)$. Also, in conducting (or superconducting) diamagnetic media, the static permittivity is required to be negative and infinite, $\varepsilon(0) \to -\infty$, due to their static conductivity [44]. Therefore, the presence of a static diamagnetic medium (of dielectric or metallic type) around a given body does not directly lead to lower integrated scattering because, while the magnetic contribution decreases, the electric one increases due to the above causality constraint. To investigate quantitatively the performance of a diamagnetic cloak we consider the relevant example of a core-shell geometry (inset of Fig. 3a), for which the integrated scattering is given by Eq. (5). In order to assess whether the diamagnetic cloak reduces the global scattering, we compute the ratio of the static polarizabilities of cloaked to uncloaked object:

$$r = \frac{\alpha_{e,cloaked} + \alpha_{m,cloaked}}{\alpha_{e,uncloaked} + \alpha_{m,uncloaked}} =$$

$$= \frac{(\varepsilon+2)(\mu+2)}{\eta^3(\varepsilon+\mu+2\varepsilon\mu-4)} \left( \begin{array}{l} \frac{\eta^3(\varepsilon-\varepsilon_s)(2\varepsilon_s+1)+(\varepsilon_s-1)(\varepsilon+2\varepsilon_s)}{2\eta^3(\varepsilon-\varepsilon_s)(\varepsilon_s-1)+(\varepsilon_s+2)(\varepsilon+2\varepsilon_s)} + \\ + \frac{\eta^3(\mu-\mu_s)(2\mu_s+1)+(\mu_s-1)(\mu+2\mu_s)}{2\eta^3(\mu-\mu_s)(\mu_s-1)+(\mu_s+2)(\mu+2\mu_s)} \end{array} \right). \quad (9)$$

Assuming the condition $\varepsilon_s = 1/\mu_s$, which is the lower bound to guarantee unitary static refractive index for causal materials, we plot in Fig. 3 the static polarizability ratio (9) in the range $r < 1$ for a dielectric diamagnetic cloak applied to a given object, varying $\mu_s$ and the volume filling factor $\eta^3$. Figs. 3a-c show the results for different objects: a dielectric sphere with $\varepsilon = 10$ and $\mu = 1$ (panel a), a magneto-dielectric sphere with $\varepsilon = 10$ and $\mu = 5$ (panel b), and a magnetic sphere with $\varepsilon = 1$ and $\mu = 10$ (panel c). We see in all the examples that, in a range of moderately high filling ratios (thin cloaks) and low permeability (strong diamagnetism), the polarizability ratio $r$ can go significantly below unity, implying that the integrated scattering of the cloaked sphere can be made lower than in the uncloaked case. We have verified that analogous results may be obtained by using diamagnetic cloaks of metallic type.

This is an important result, as it proves that properly tailored diamagnetic cloaks can indeed lower the scattering on a global scale. The region where $r < 1$ and the value of minimum $r$ depend on the object to be cloaked. In general, if its static magnetic permeability is larger, we may be able to obtain a lower $r$, since the diamagnetic cloak can directly compensate the contribution to the global scattering provided by the static magnetic polarizability of the object. As an example, we show in Fig. 3d, the evolution of the boundary $r = 1$ varying the core permeability. As expected, the region of lower global scattering, delimited by the curves in Fig.

3d, broadens when $\mu$ increases. Even for nonmagnetic objects (lowest curve in the figure, corresponding to the case in Fig. 3a) a good global scattering reduction can be obtained by using suitably designed diamagnetic cloaks.

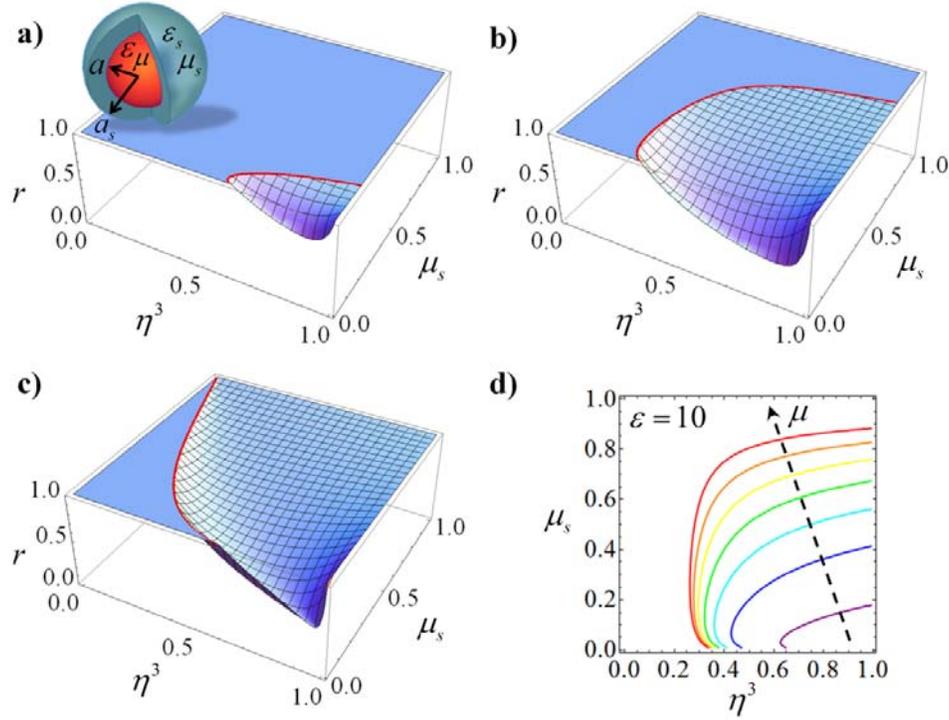

*Figure 3 – A spherical object covered by a dielectric diamagnetic cloak, as depicted in the inset of panel (a), with $0 < \mu_s < 1$ and $\varepsilon_s = 1/\mu_s$. (a), (b), (c) Static polarizability ratio $r$ between uncloaked and cloaked scenarios, calculated through Eq. (9), as a function of the filling factor $\eta^3$ and the relative static permeability $\mu_s$ of the cloak. Three spherical objects with different material properties are considered: (a) $\varepsilon = 10$, $\mu = 1$, (b) $\varepsilon = 10$, $\mu = 5$ and (c) $\varepsilon = 1$, $\mu = 10$. In each case, only the region where the cloak reduces the integrated scattering, i.e., $r < 1$, is shown. The red thick curves indicate where $r = 1$, delimiting the region of interest. (d) This boundary, $r = 1$, is evaluated varying the object permeability $\mu$, for $\varepsilon = 10$. From bottom to*

*top, $\mu$ changes from 1 to 7 with unitary steps. The region in which the diamagnetic cloak is effective in reducing the integrated scattering is below each curve.*

All these examples confirm that a rather strong diamagnetism is required to attain a noticeable reduction in integrated scattering. All conductors, e.g., silver and bismuth [32],[36], exhibit moderate diamagnetism due to eddy currents (metallic-type diamagnetism, or Landau diamagnetism [32]), as discussed in Sec. II, however, this effect is still too weak for our purposes (and it is often partially or completely overcome by the paramagnetic properties of the material, as for aluminum). Dielectric materials also have generally weak diamagnetic properties (dielectric-type diamagnetism, or Langevin diamagnetism [33]), while some metamaterials may be able to achieve strong static diamagnetic response [47].

It is evident from the results in Fig. 3 that the maximum global cloaking effect is obtained when $\mu_s \to 0$, which requires $|\varepsilon_s| \to \infty$ to respect the causality constraint on the static refractive index, and a filling ratio $\eta \to 1$ to allow the wave to penetrate the cloak. In Sec. II we have discussed how these constitutive properties are typical of a perfect conductor, in which Landau diamagnetism is ideally strong. Although PEC materials don't exist in nature, superconductors may well approximate their behavior [33],[45]. Superconducting media have been already proposed in the context of magnetostatic cloaking based on scattering cancellation [47]-[48], and our results suggest that they may also be capable of drastically reducing the global integrated scattering of a given object over all frequencies.

In order to further investigate the global performance of diamagnetic cloaks as a function of the material properties of the object, we plot in Fig. 4a the evolution of the minimum polarizability ratio $r$ varying the static permittivity and permeability of the object to be cloaked, assuming that the dielectric diamagnetic cloak is characterized by $\mu_s = 0.01$ and $\varepsilon_s = 1/\mu_s = 100$. Each point of the plot represents a different object, cloaked by the mentioned diamagnetic material, with optimal thickness to select the minimum attainable value of $r$. We see in Fig. 4a that the global cloaking performance improves monotonically when the constitutive parameters of the object increase in value, consistent with what observed in Fig. 3 for the considered example. It is easy to show from Eq. (9) that, for an object with very large constitutive parameters, the optimal cloak is composed of an ultrathin shell made of ideal diamagnetic material, and taking the limit of (9) for $\mu_s \to 0$, $|\varepsilon_s| \to \infty$ and $\eta \to 1$, we find a minimum polarizability ratio of $1/4$, which corresponds to a 75% reduction of the integrated scattering from the object. Our theory proves that this value is the theoretical maximum for global scattering reduction.

As a specific example, consider the case of a homogenous sphere with $\varepsilon = 10$ and $\mu = 5$, covered by a strongly diamagnetic cloak with $\mu_s = 0.01$ and $\varepsilon_s = 1/\mu_s = 100$, which corresponds to the red circle in Fig. 4a. As shown in Fig. 4b, in this case the minimum polarizability ratio is 0.43, obtained for $\eta^3 = 0.874$ (corresponding to a cloak radius $a_s = 1.046a$). Assuming constant material properties with frequency (which is consistent with causality, since the refractive index is always unitary [46]), we plot in Fig. 4c the SCS of the uncloaked (black line) and cloaked (blue) spheres over a broad wavelength range. Covering the object with a strongly diamagnetic film indeed drastically reduces its global scattering, yielding $I_{scat} = 2.05 \cdot 10^{-2} \, \mu m^3$ for the uncloaked sphere and $I_{scat} = 9.22 \cdot 10^{-3} \, \mu m^3$ when the cloak is applied. These results confirm that the designed diamagnetic cloak can reduce the integrated scattering by 2.2 times. In most (yet not all) frequency ranges the cloaked object is indeed less visible than the uncloaked case.

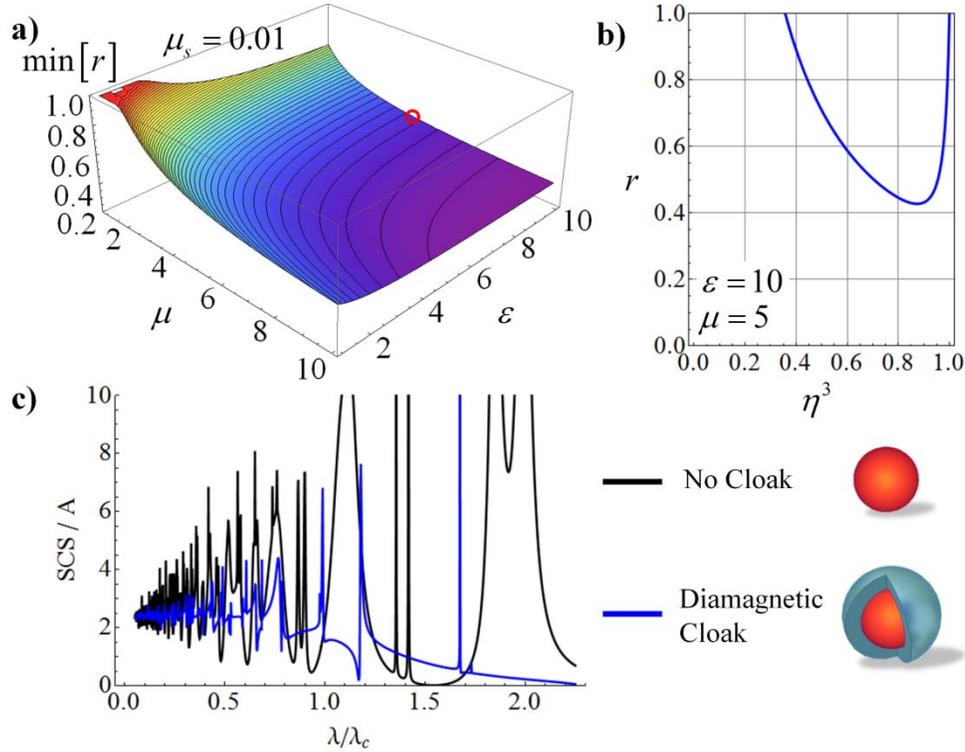

Figure 4 – (a) Evolution of the minimum attainable polarizability ratio $r$ varying the material properties of the spherical object on which the cloak is applied. The diamagnetic cloak considered here has $\mu_s = 0.01$ and $\varepsilon_s = 1/\mu_s = 100$ and, for each point of the plot, the optimal filling factor $\eta^3$ to minimize $r$. (b) Polarizability ratio as a function of $\eta^3$, for the same cloak material considered in (a) and an inner object with $\varepsilon = 10$ and $\mu = 5$, corresponding to the red circle in (a). A minimum ratio $r = 0.43$ is achieved for $\eta^3 = 0.874$. (c) Corresponding normalized SCS over a broad wavelength range for uncloaked (black curve) and cloaked (blue) cases in the optimal design as in (b).

The results presented in this section demonstrate that diamagnetic media, and superconductors in particular, are the only passive and linear materials that, when applied around a given body, can lower its global scattering properties. Notably, in the case of a core-shell geometry the integrated scattering can be reduced up to one fourth, depending on the object to be cloaked.

## IV. CONCLUSIONS

The findings presented in this paper shed new light on the global scattering behavior of invisibility cloaks. We have found that any passive and linear cloak covering a given object necessarily increases its scattering integrated over all wavelengths, unless the cloaking material has strong static diamagnetic properties and a suitably tailored design. This implies that all currently proposed cloaking techniques, while being effective at lowering the SCS over a narrow frequency window, necessarily scatter significant power at other frequencies and, therefore, they may be defeated (and detected more easily than the original uncloaked objects) when excited with short pulses. To overcome this limitation, we have proposed optimal designs for global cloaks that can suppress the scattering integrated over all wavelengths by using suitably tailored thin diamagnetic/superconducting shells, and we envision that these general bounds may be further relaxed by considering active or nonlinear cloaks [49].

Our results may be extended to combine local and global cloaking techniques, to drastically suppress the scattering around the desired frequency, as well as reduce the scattered energy over the entire spectrum. For instance, by patterning a superconducting thin film [50] a mantle cloaking effect may be achieved at the desired frequency, while noticeably lowering the global integrated scattering at the same time. The use of superconductors may also provide the possibility of dynamically switching the cloaking device by changing the temperature of the superconducting cloak around its critical point. We believe that the development of the concepts and ideas proposed in this paper may open new and exciting research lines in the field of cloaking and invisibility, as well as allow a more quantitative assessment of the global performance of cloaking techniques, and their overall detectability. Since a similar version of the integrated scattering sum rule holds for acoustic waves [51], we expect that the present bounds apply also to acoustic cloaking.

## ACKNOWLEDGEMENTS

This work was partially supported by the DTRA YIP award No. HDTRA1-12-1-0022, and the AFOSR YIP award No. FA9550-11-1-0009.